\documentclass[structabstract]{aa}  
\usepackage{graphicx}
\usepackage{txfonts}
\usepackage{natbib}
\begin{document}
\title{Carbon in different phases ([CII], [CI], and CO) in infrared
  dark clouds: Cloud formation signatures and carbon gas fractions
  \thanks{Based on observations carried out with Herschel, Sofia, APEX,
    and the IRAM 30\,m telescope. The data are available in electronic
    form at the CDS via anonymous ftp to cdsarc.u-strasbg.fr
    (130.79.128.5) or via
    http://cdsweb.u-strasbg.fr/cgi-bin/qcat?J/A+A/}.}

%   \subtitle{I. Overviewing the $\kappa$-mechanism}

   \author{H.~Beuther
          \inst{1}
          \and
          S.E.~Ragan
          \inst{1}
          \and
          V.~Ossenkopf
           \inst{2}
          \and
          S.~Glover
           \inst{3}
          \and
          Th.~Henning
           \inst{1}
          \and
          H.~Linz
          \inst{1}
          \and
          M.~Nielbock
           \inst{1}
          \and
          O.~Krause
           \inst{1}
          \and
          J.~Stutzki
           \inst{2}
          \and
          P.~Schilke
           \inst{2}
          \and
          R.~G\"usten
           \inst{4}
            }
   \institute{$^1$ Max Planck Institute for Astronomy, K\"onigstuhl 17,
              69117 Heidelberg, Germany, \email{name@mpia.de}\\
              $^2$ I. Physikalisches Institut, University of Cologne, Z\"ulpicher Strasse 77, 50937 K\"oln, Germany\\
             $^3$ Center for Astronomy, Institute for Theoretical Astrophysics, Albert-\"Uberle Strasse 2, 69120 Heidelberg, Germnany\\
             $^4$ Max Planck Institute for Radioastronomy, Auf dem H\"ugel 69, 53121 Bonn, Germany}

   \date{Version of \today}

%   \abstract{}
% \abstract{}{}{}{}{} 
% 5 {} token are mandatory
\abstract
{How molecular clouds form out of the atomic phase and what
  the relative fractions of carbon are in the ionized, atomic, and
  molecular phase are questions at the heart of cloud and star
  formation.}
{We want to understand the kinematic processes of gas flows during the
  formation of molecular clouds. In addition to that, we aim at
  determining the abundance ratios of carbon in its various gas
  phases from the ionized to the molecular form.}
{Using multiple observatories from Herschel and SOFIA to APEX and the
  IRAM 30\,m telescope, we mapped the ionized and atomic carbon as well as carbon monoxide ([CII] at 1900\,GHz, [CI] at 492\,GHz, and
      C$^{18}$O(2--1) at 220\,GHz) at high spatial resolution
  ($12''-25''$) in four young massive infrared dark clouds (IRDCs).}
{The three carbon phases were successfully mapped in all four regions,
  only in one source does the [CII] line remain a non-detection. With
  these data, we dissect the spatial and kinematic structure of the
  four IRDCs and determine the abundances of gas phase carbon in
      its ionized, atomic, and most abundant molecular form (CO).
  Both the molecular and atomic phases trace the dense structures
  well, with [CI] also tracing material at lower column densities.
  [CII] exhibits diverse morphologies in our sample from compact to
  diffuse structures, probing the cloud environment.  In at least two
  out of the four regions, we find kinematic signatures strongly
  indicating that the dense gas filaments have formed out of a
  dynamically active and turbulent atomic and molecular cloud, potentially
  from converging gas flows. The atomic carbon-to-CO gas mass
  ratios are low between 7\% and 12\% with the lowest values found
  toward the most quiescent region.  In the three regions where [CII]
  is detected, its mass is always higher by a factor of a few than
  that of the atomic carbon. While the ionized carbon emission
      depends on the radiation field, we also find additional
      signatures that indicate that other processes, for example, energetic gas
      flows can contribute to the [CII] excitation as well.}
{Combining high-resolution maps in the different carbon phases reveals
  the dynamic interplay of the various phases of the interstellar
  medium during cloud formation. Extending these studies to more
  evolved stages and combining the observations with molecular cloud
  formation simulations including the chemistry and radiative
  transfer will significantly improve our understanding of the
  general interstellar medium, cloud and star formation processes.}
\keywords{Stars: formation -- Stars: early-type -- Stars: individual:
  G11.11, G48.66, IRDC\,18223, IRDC\,18454 -- Stars: massive -- ISM: clouds -- ISM: structure}

\titlerunning{Different phases of carbon in IRDCs}

\maketitle

\section{Introduction}
\label{intro}

Carbon is one of the most important species in the interstellar medium
(ISM), both in terms of its relevance for the physical and chemical state
of the ISM, as well as because of its diagnostic power (e.g.,
\citealt{henning1998}). Because of its high abundance, carbon is one
of the major coolants of the ISM, in its atomic and ionized form for
the low-extinction components and in its most abundant molecular form
(CO) in high extinction and dense molecular clouds.  There are two
main pathways linking the sequence C$^+$/C$^0$/CO: During cloud
formation, the originally diffuse and ionized medium recombines and
becomes partly neutral (observable in lines of ionized C$^+$ ([CII])
and neutral carbon C$^0$ ([CI])), and then with increasing densities
forms molecular H$_2$ and CO (e.g., \citealt{bergin2004,glover2010}).
In contrast, if exposed to strong UV radiation, CO can dissociate to
form atomic and ionized carbon (e.g.,
\citealt{koester1994,roellig2006}).

As atomic hydrogen reacts to form H$_2$ during cloud formation (e.g.,
\citealt{bergin2004,heitsch2008}), the carbon follows the
C$^+\rightarrow$C$^0\rightarrow$CO route during the formation of
infrared dark clouds (IRDCs). At early evolutionary stages, C$^+$ is
likely to be one of the best tracers for the so-called ``dark H$_2$''
because due to self-shielding H$_2$ forms earlier than CO (e.g.,
\citealt{langer2010,glover2011}).  During these processes, signatures
of streaming motions may be imprinted on the different carbon lines,
allowing us to investigate the cloud formation processes (e.g.,
\citealt{glover2010,shetty2011,clark2012}).
%For example, in dynamical
%cloud formation models one expects [CII] to have a larger line width
%than CO, since the former should primarily be tracing the converging
%gas streams, while the latter comes mainly from the dense cloud formed
%at the stagnation point of the convergent flow.

At the center of IRDCs, massive stars form rapidly, exciting the
surrounding gas and producing the different forms of carbon via the
opposite CO$\rightarrow$C$^0\rightarrow$C$^+$ route.  Photon-dominated
region (PDR) models give predictions about the line properties in this
scenario (e.g., \citealt{tielens1985a,cubick2008}). Furthermore, the
youngest and most embedded protostars develop jets and outflows that
form cavities that act as valves releasing UV radiation into the
ambient gas (e.g., \citealt{kuiper2010,cyganowski2008}). Thus,
feedback from the youngest protostars affects the relative abundances
of molecular CO and atomic and ionized carbon also in the most
deeply embedded regions.

Early work has shown that [CII] and [CI] emission is far more extended
through molecular clouds than was initially anticipated with
layered-structure cloud models.  This implied that molecular clouds
are clumpy and that the radiation can penetrate the clouds deeply
(e.g., \citealt{stutzki1988,herrmann1997,kramer2004}). However,
because of the poor spatial resolution available at those times, in
particular for C$^+$, there was no chance to spatially resolve the
substructures, and to disentangle the excitation mechanisms of the
different gas components.  For example, is the [CII] and [CI] emission
deep inside the clouds mainly due to external radiation that
penetrates deeply through the clumpy medium, or do internal heating
sources from embedded stars contribute significantly to the formation
of C$^+$ and C$^0$ (e.g., \citealt{ossenkopf2011})?  Topics like that
can be addressed particularly well for IRDCs because, while they are
at early stages of cloud evolution, many already contain internal
heating sources that excite the immediate environment (e.g.,
\citealt{rathborne2006,beuther2010b,henning2010,wang2014}). This
information is also necessary for the proper modeling of the structure
and evolution of molecular clouds, since any interior warm gas
provides thermal support and deep penetration of UV photons enhances
the importance of magnetic support.

To disentangle the different carbon components, it is essential to map
the respective regions at high spatial resolution, and to spectrally
resolve the line emission.  The detailed comparison of the line
profiles and their small scale variation between [CII]/[CI]/CO allows
us to separate clumps/filaments overlapping along the line of sight.
While mapping is important to spatially differentiate the origin of
the ionized/atomic/molecular carbon phases, spectral line shape
information can be directly compared to cloud formation and PDR model
predictions (e.g., \citealt{cubick2008,glover2010}), as well as to the
impact of outflows from the embedded protostars.  Mapping at high
spectral resolution [CII]/[CI]/CO with Herschel/SOFIA/APEX/IRAM30\,m
(the Herschel satellite mission, the Stratospheric Observatory For
Infrared Astronomy, The Atacama Pathfinder Experiment, and the 30\,m
telescope of the Institut de Radioastronomie Millimetrique),
respectively, can result in spatial resolution elements between $11''$
and $15''$, corresponding to linear scales below 0.2\,pc at typical
distances of IRDCs of 3\,kpc. This matches the physical processes of
cloud formation, fragmentation and feedback extremely well, e.g., the
Jeans fragmentation scale at typical densities ($\sim
10^3$\,cm$^{-3}$) is on that order.

To address these aspects, we conducted a concerted effort to study the
gas phase carbon budget in young molecular clouds by observing and
analyzing a well selected sample of four IRDCs in [CII] at 158\,$\mu$m
with Herschel and SOFIA, in [CI] with APEX and in
$^{13}$CO/C$^{18}$O(2--1) with the IRAM 30\,m telescope. This combined
mapping approach is complementary to the GOTC$^+$ Herschel key project
(Galactic Observations of Terahertz C$^+$,
\citealt{langer2010,pineda2013}) that focused on single pointings
toward a large sample of Galactic sources. Now mapping the regions
allows us to separate the different cloud components and their
associated carbon constituents properly from low to high extinction.
These data provide us with the column densities, the spatial structure
and the kinematic/dynamic properties of the different states of carbon
in the ISM. For all regions we have complementary PACS/SPIRE continuum
data (EPOS Herschel GT project, \citealt{ragan2012b}) as well as
APEX/LABOCA 870\,$\mu$m continuum observations \citep{schuller2009}.

\begin{figure}[htb] 
\includegraphics[width=0.49\textwidth]{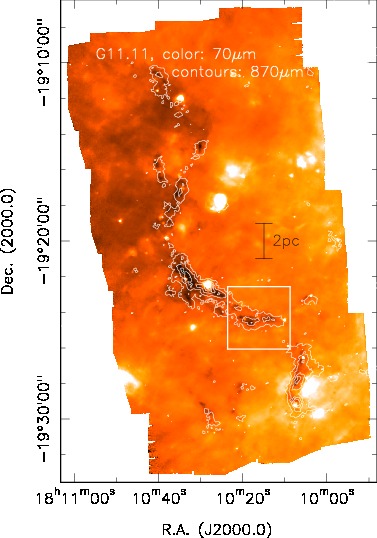}
\caption{G11.11: Large-scale Herschel/PACS 70\,$\mu$m image of the
  G11.11 region with 870\,$\mu$m ATLASGAL contours starting at 200 and continuing in
  300\,mJy\,beam$^{-1}$ steps \citep{henning2010}. The white box outlines the region
  of our carbon observations shown in Fig.~\ref{g11_overlays}.}
\label{g11_large}
\end{figure}

\section{The Sample}
\label{sec_sample}

We selected four infrared dark clouds that exhibit a variety of
environments.  While G11.11 (also know as The Snake, e.g.,
\citealt{henning2010,kainulainen2013}) and G48.66 (e.g.,
\citealt{ossenkopf2011,pitann2013}) are rather isolated IRDCs (G48.66
shows slightly more star formation activity), IRDC\,18223 is an
infrared dark filament that has already formed a high-mass
protostellar object with $\sim 10^4$\,L$_{\odot}$ at one end (e.g.,
\citealt{beuther2010b}). Finally, IRDC\,18454 is also a starless dark
cloud, however in the very close environment of the Galactic
mini-starburst W43 \citep{beuther2013a}. All these clouds are part of
the Herschel key project EPOS with a wealth of complimentary data
\citep{ragan2012b}. Table \ref{sample} presents the basic sample
parameters.

\begin{table}[htb]
\caption{The Sample}
\begin{tabular}{lrrrr}
\hline \hline
Name & R.A. & Dec. & D & $v_{\rm{lsr}}$ \\
     & (J2000.0) & (J2000.0)  &  (kpc) & (km\,s$^{-1}$) \\
\hline
G11.11    & 18:10:16.00 & -19:24:22.0 & 3.4 & 29.2  \\
G48.66    & 19:21:48.00 &  13:49:06.0 & 2.6 & 34.0  \\
IRDC18223 & 18:25:08.46 & -12:45:05.0 & 3.5 & 45.5  \\
IRDC18454 & 18:47:59.20 & -01:54:05.0 & 5.5$^1$ & 52.8/100.2$^2$  \\
\hline \hline
\end{tabular}
~\\
{\footnotesize Most distances and $v_{\rm{lsr}}$ are taken from \citet{ragan2012b}.\\
$^1$ \citet{zhang2014}, $^2$ \citet{beuther2007g}}
\label{sample}
\end{table}

\begin{figure*}[htb] 
\includegraphics[width=0.99\textwidth]{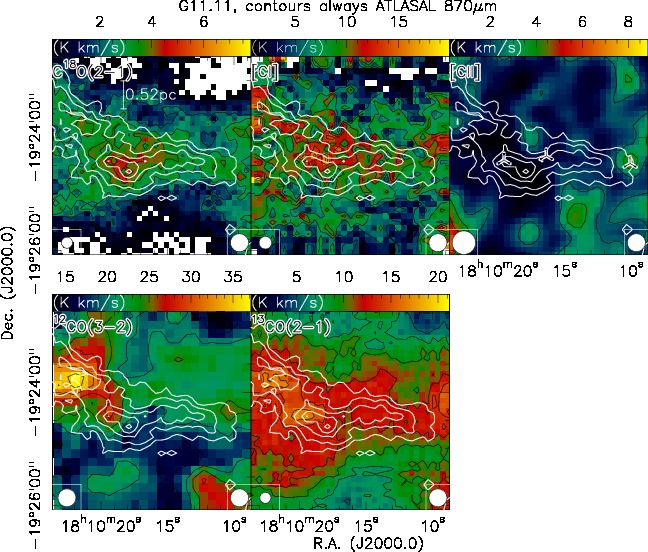}
\caption{G11.11: The color scale shows integrated intensity images of
  the transitions marked above each panel. Integration regimes
  are [27,33]\,km\,s$^{-1}$, [27,33]\,km\,s$^{-1}$,
  [29,31]\,km\,s$^{-1}$, [24,38]\,km\,s$^{-1}$ and
  [24,36]\,km\,s$^{-1}$ for C$^{18}$O(2-1), [CI], [CII],
  $^{12}$CO(3-2) and $^{13}$CO(2-1), respectively. Except for [CII],
  the corresponding black contours start from 15\% and continue in
  10\% steps of the peak emission in each map. The peak values are
  7.8, 19.9, 37.1 and 21.2\,K\,km\,s$^{-1}$, respectively.  The [CII]
  contours start at 3$\sigma$ and continue in 1$\sigma$ steps of
  0.8\,K\,km\,s$^{-1}$. The white contours always show the ATLASGAL
  870\,$\mu$m continuum image starting at a $3\sigma$ level of
  0.15\,Jy\,beam$^{-1}$ and continue in $4\sigma$ steps. The
  bottom-left of each panel shows the beam of the line data whereas
  the bottom-right shows the continuum beam size. The three markers in
  the [CII] panel show the positions of 70\,$\mu$m sources, and the
  top-left panel also shows a linear scale-bar.}
\label{g11_overlays}
\end{figure*}

\section{Observations} 
\label{obs}

\begin{table}[htb]
\caption{Observed spectral lines}
\begin{tabular}{lrr}
\hline \hline
Freq. & Transitions & $E_{\rm{u}}/k$ \\
(GHz) &       &  (K) \\
\hline
1900.5372 & [CII]($^2P_{3/2}-^2P_{1/2}$) & 91.2 \\
492.1607 & [CI]($^3P_1-^3P_0$) & 23.6  \\
219.5603 & C$^{18}$O(2--1) & 15.8 \\
220.3987 & $^{13}$CO(2--1) & 15.9 \\
345.7960 & $^{12}$CO(3--2) & 33.2 \\
\hline \hline
\end{tabular}
\label{linelist}
\end{table}

\begin{figure}[htb] 
\includegraphics[width=0.49\textwidth]{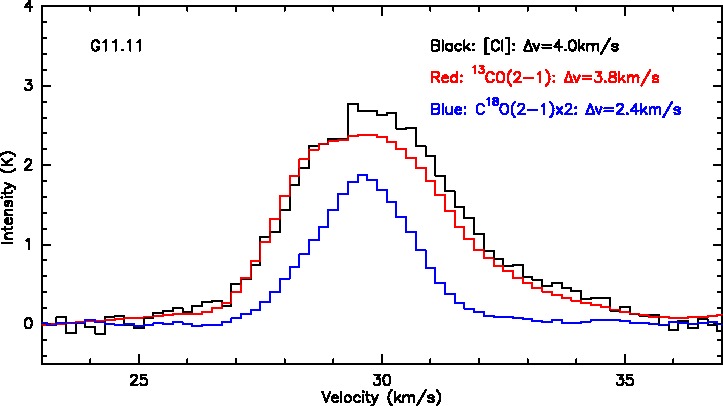}
\caption{G11.11: C$^{18}$O(2-1), $^{13}$CO(2--1) and [CI] spectra
  averaged over the whole area of emission shown in Figure
  \ref{g11_overlays}.}
\label{g11_spectra}
\end{figure}

\begin{figure*}[htb] 
\includegraphics[width=0.95\textwidth]{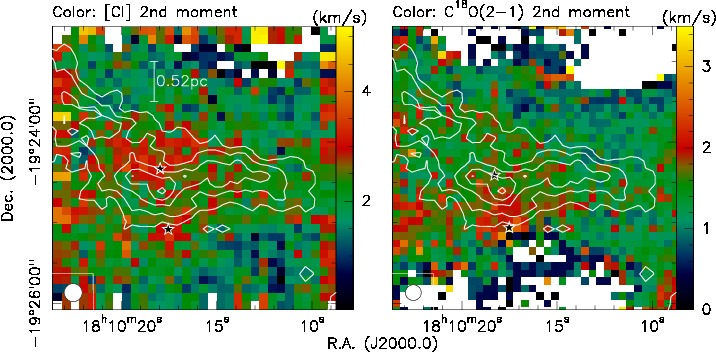}
\caption{G11.11: The color scales show the 2nd moment maps (intensity
  weighted line widths) of [CI] and C$^{18}$O(2-1) in the left and
  right panel, respectively. The contours present the ATLASGAL
  870\,$\mu$m emission with the same contours as in
  Fig.~\ref{g11_overlays}. The bottom-left of each panel shows the
  beam of the line data, and the left panel also presents a linear
  scale-bar. The five-pointed stars mark the positions toward
    which the spectra in Fig.~\ref{g11_spectra_peaks} are extracted.}
\label{g11_mom2}
\end{figure*}

The different spectral lines shown in Table \ref{linelist} were
observed with various instruments, the ionized carbon [CII] line with
Herschel/HIFI
(\citealt{A&ASpecialIssue-HERSCHEL,A&ASpecialIssue-HIFI}, for G11.11,
G48.66, IRDC18454) and SOFIA/GREAT (\citealt{young2012,heyminck2012},
for IRDC\,18223), the atomic carbon [CI]$(^3P_1-^3P_0)$ and
$^{12}$CO(3--2) with APEX, and the C$^{18}$O(2--1) and $^{13}$CO(2--1)
with the IRAM\,30\,m telescope.  All data are calibrated to
$T_{\rm{mb}}$ in Kelvin. The final spectral resolution $\Delta v$, the
$1\sigma$ rms in a line-free channel and the spatial resolution
$\theta$ for all data can be found in Table \ref{obs_par}.

\begin{table*}[htb]
\caption{Observing parameters}
\begin{center}
\begin{tabular}{l|rrr|rrr|rrr}
\hline \hline
Source & $\Delta v$ & $1\sigma$ & $\theta$ & $\Delta v$ & $1\sigma$ & $\theta$ & $\Delta v$ & $1\sigma$ & $\theta$  \\
     & (km\,s$^{-1}$) & (K) & ($''$) & (km\,s$^{-1}$) & (K) & ($''$) & (km\,s$^{-1}$) & (K) & ($''$) \\
 & \multicolumn{3}{c}{C$^{18}$O(2--1)} & \multicolumn{3}{|c|}{[CI]}  & \multicolumn{3}{|c|}{[CII]$^1$} \\ 
\hline
G11.11 & 0.4 & 0.16 & 12 & 0.4 & 0.6 & 13.5 & 0.08 & 0.32 & 25 \\
G48.66 & 0.27 & 0.25 & 12 & 0.2 & 1.0 & 13.5 & 0.08 & 0.31 & 20 \\
IRDC18223 & 0.27 & 0.2 & 12 & 0.3 & 0.6 & 13.5 & 0.2 & 0.28 & 20 \\
IRDC18454 & 1.5 & 0.13 & 12 & 0.2 & 0.9 & 13.5 & 0.08 & 0.38 & 20 \\
\hline \hline
\end{tabular}
~\\
{\footnotesize $\Delta v$, $1\sigma$ and $\theta$ are the spectral resolution, 1$sigma$ rms and beam size, respectively, for the three transitions.\\
$^1$ The [CII] data from G11.11, G48.66 and IRDC\,18454 are from Herschel, the [CII] data for IRDC\,18223 are from SOFIA.}
\end{center}
\label{obs_par}
\end{table*}

The Herschel/HIFI [CII] data for G11.11, G48.66 and IRDC18454 were
observed in one guaranteed time (obids 1342250966, 1342250967) and one
open time project (obsids 1342270620, 1342270621, 1342270624,
1342270625).  The approximate map sizes were in all cases $\sim
3'\times 3'$, and each map was observed twice at scanning angles of 0
and 90\,deg. Data reduction was conducted within HIPE version 10.3 and
beam and forward efficiencies of 0.69 and 0.96 were applied
\citep{roelfsema2012}. The observations were taken against the
internal cold load as reference and included the observation of an
additional OFF position at an offset of R.A.\, $+5'$ and Dec.\, $-5'$
from the center of the map. As we found clear contamination of the OFF
position, we only subtracted the load reference from the data, and
removed the remaining standing wave pattern by the HifiFitFringe
pipeline task which resulted in good baselines for our spectra.  The
spectra were then exported to GILDAS format for final processing and
imaging. For G11.11 we used only the V-polarization because that had
less standing wave problems. For the other two sources, we used both
polarizations. To increase the signal-to-noise ratio, we degraded the
native spatial resolution of $\sim 12''$ to $20''-25''$ (Table
\ref{obs_par}).

The corresponding [CII] data for IRDC\,18223 were observed in fall
2013 with GREAT on SOFIA and a beam efficiency of 0.67
\citep{heyminck2012}.  These data were smoothed to $20''$ resolution
to increase the signal-to-noise ratio as well.

The atomic carbon [CI] data were all obtained with the APEX
observatory and the FLASH receiver. The maps have similar sizes as the
[CII] maps and were observed in on-the-fly mode at a spatial
resolution of $\sim 13.5''$ and a native spectral resolution of
0.05\,km\,s$^{-1}$. We re-binned the spectra to improve the signal to
noise, and the finally used effective spectral resolution $\Delta v$,
$1\sigma$ and $\theta$ can be found in Table \ref{obs_par}.  The FLASH
receiver observed simultaneously the $^{12}$CO(3--2) line, but we show
that only for G11.11 as an example because our analysis relies on the
C$^{18}$O(2--1) data due to its lower optical depth.

Finally, the C$^{18}$O(2--1) spectra were all observed during
different runs with the IRAM\,30\,m telescope. The maps were taken in
the on-the-fly mode, and the spatial resolution is $\sim 12''$. The
data for G11.11, G48.66 and IRDC\,18223 were observed as pooled
observations in winter 2012. The observations for IRDC\,18454 were
part of an IRAM large program (PI F.~Motte) and have already been
published in \citet{carlhoff2013}. The IRAM 30\,m telescope also
observed simultaneously the $^{13}$CO(2--1) line, but we again only
show it for G11.11 as an example and rely for the analysis on the
C$^{18}$O(2--1) data. The corresponding 870\,$\mu$m continuum data
were taken from the ATLASGAL survey \citep{schuller2009}.

\begin{figure}[htb] 
\includegraphics[width=0.49\textwidth]{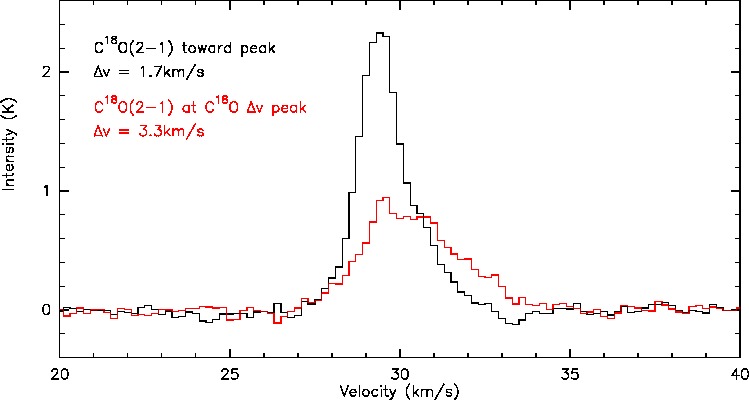}
\includegraphics[width=0.49\textwidth]{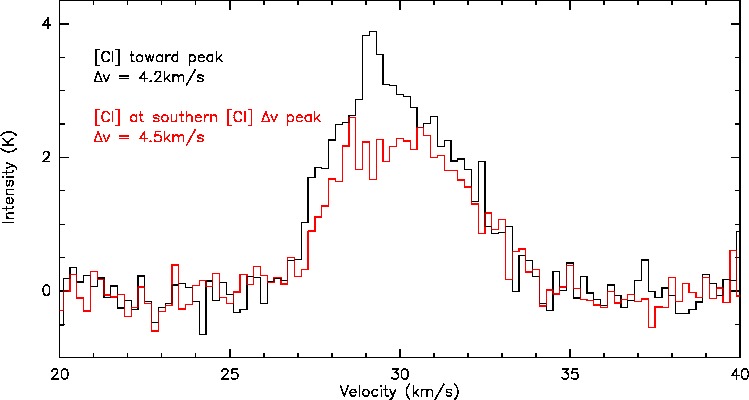}
\caption{G11.11: C$^{18}$O(2-1) (top-panel) and [CI] spectra (bottom
  panel) at selected positions as marked in the panel and in
    Figure \ref{g11_mom2}.}
\label{g11_spectra_peaks}
\end{figure}

\section{Results}

\subsection{Morphologies and structures}

Since our data are the first maps in the different gas carbon phases,
we start with a morphological and structural analysis of the data
toward the four IRDCs.  

\subsubsection{G11.11 also know as The Snake:} 

\begin{figure}[htb] 
\includegraphics[width=0.49\textwidth]{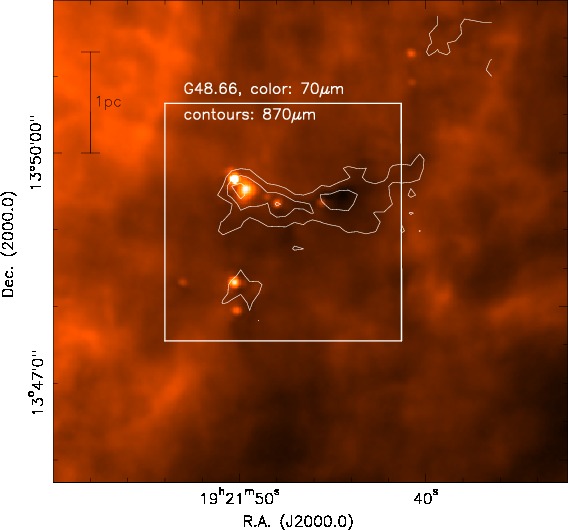}
\caption{G48.66: Large-scale Herschel/PACS 70\,$\mu$m image of the
  G48.66 region with 870\,$\mu$m ATLASGAL contours in $3\sigma$ steps of
  150\,mJy\,beam$^{-1}$. The white box outlines the region of our
  carbon observations shown in Fig.~\ref{g48_overlays}
  \citep{pitann2013}.}
\label{g48_large}
\end{figure}

The IRDC G11.11 encompasses a much larger area than shown in
Fig.~\ref{g11_large}. Here we focus on the $3'\times 3'$ region
outlined by the white box. That subregion is a particularly quiescent
part of the G11.11 region, hosting only very few weak mid-infrared
sources as well as a dark core without any mid- or far-infrared
counterpart \citep{henning2010,ragan2012b}. Figure \ref{g11_overlays}
presents the different carbon datasets we obtained for this project.
The 870\,$\mu$m submm continuum data tracing the cold dust emission
are shown in contours as reference frame on all plots. The first and
most obvious outcome for G11.11 is that the molecular C$^{18}$O
and atomic carbon emission do trace the high-column density gas probed
by the 870\,$\mu$m continuum emission well. While the C$^{18}$O(2--1)
emission follows closely the dust continuum, the atomic [CI] emission
appears a bit more extended. In contrast to these two tracers, the
ionized carbon [CII] emission remains a non-detection in our data.
Towards the dust continuum, we do not detect any [CII] at all, but a
few very weak features are tentatively identified at the cloud edges.
However, none of these integrated emission edge features are above
$6\sigma$, and analyzing individual spectra at these position, we
cannot confirm them either.  Hence, at the given sensitivity, our
[CII] observations toward G11.11 are a non-detection, even after
averaging all spectra over the entire region.

Figure \ref{g11_overlays} also shows the $^{12}$CO(3--2) emission from
APEX and the $^{13}$CO(2--1) emission from the IRAM30\,m telescope.
However, both trace the dense gas emission outlined by the dust
continuum worse than the C$^{18}$O(2--1). This is mainly due to the
larger optical depth of these isotopologues. For the remainder of the
analysis, we therefore focus on the C$^{18}$O(2--1) data because of
its low optical depth.

\begin{figure*}[htb] 
\includegraphics[width=0.99\textwidth]{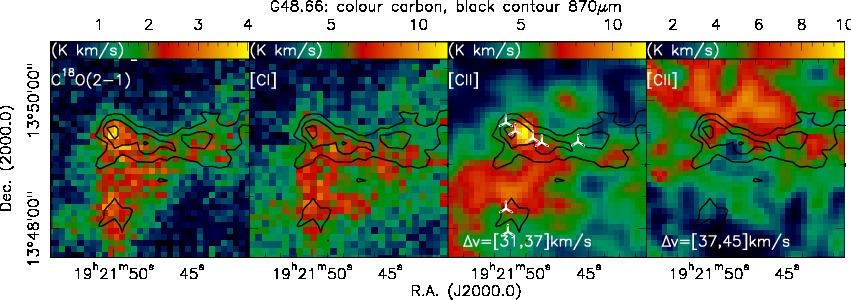}
\caption{G48.66: The color scale shows from left to right the emission
  from C$^{18}$O(2--1), [CI] and [CII], respectively. The integration
  ranges for C$^{18}$O(2--1) and [CI] are [32.0;35.0] and
  [32.0;35.5]\,km\,s$^{-1}$.  The integration ranges for [CII] are
  marked in the 2 right panels. The contours always show the ATLASGAL
  870\,$\mu$m emission starting at $3\sigma$ level of
  150\,mJy\,beam$^{-1}$ and continuing in $3\sigma$ steps.  The white
  markers in the 3rd panel mark the positions of 70\,$\mu$m sources.}
\label{g48_overlays}
\end{figure*}

Figure \ref{g11_spectra} presents the averaged spectra over the entire
region of emission in Figure \ref{g11_overlays} for C$^{18}$O(2--1),
$^{13}$CO(2--1) and [CI]. While the line shape does vary significantly
between the different species, the line width measured as full width
half maximum is similar for [CI] and $^{13}$CO(2--1) ($\Delta v=4.0$
and 3.8\,km\,s$^{-1}$, respectively), whereas it is narrower for
C$^{18}$O(2--1) ($\Delta v=2.4$\,km\,s$^{-1}$). Since $^{13}$CO and
[CI] trace larger volumes than C$^{18}$O, this difference is
approximately consistent with Larson's line width-size relation
\citep{larson1981,stahler2005}.  Although the averaged line width over
the whole cloud of [CI] and $^{13}$CO(2--1) are similar, their spatial
structure varies considerably.

For a closer comparison of the C$^{18}$O(2--1) and [CI] line width
distribution, Figure \ref{g11_mom2} presents 2nd moment maps
(intensity-weighted line widths) of both lines. Interestingly, the
[CI] line width peaks close to the main submm continuum sources and
hence resembles the [CI] integrated intensity image in Figure
\ref{g11_overlays}. Together with the non-Gaussian profiles, this is
consistent with a macro-turbulent picture, where individual, partially
optically thick clumps are distributed over some turbulent velocity
interval. In contrast to that, the C$^{18}$O(2--1) 2nd moment map
shows a relatively smooth velocity distribution over most of the clump
with only an increase in line width south of the main submm continuum
peak. The same position shows also an increased line width in the [CI]
emission. Figure \ref{g11_spectra_peaks} present the C$^{18}$O(2--1)
and [CI] spectra extracted toward the respective integrated intensity
as well as line width peak positions, and the same trend is visible
there. While the [CI] line widths between both positions does not vary
much, for C$^{18}$O(2--1) we find a unique line width increase toward
that southern position at the edge of cloud. Interestingly, Ragan et
al.~(in prep.) find toward the close-by submm continuum peak position
(Fig.~\ref{g11_mom2}) multiple N$_2$H$^+$ velocity components in
high-spatial-resolution Plateau de Bure Interferometer (PdBI) data.
These multiple N$_2$H$^+$ spectral features are consistent with global
collapse of massive gas clumps as modeled by \citet{smith2013}.
Although the positions of our enhanced C$^{18}$O and [CI] line width
do not exactly coincide with the multiple N$_2$H$^+$ spectra (offset
$\sim 25''$), the close spatial association may indicate that both
features could trace different parts of the same global collapse of
this star-forming region.

\subsubsection{G48.66:} 

The IRDC G48.66 is a prominent massive dark cloud in the projected
vicinity of W51 (distance $\sim$5.4\,kpc, \citealt{sato2010}),
however at closer distance of 2.6\,kpc (e.g.,
\citealt{ormel2005,vanderwiel2008}). A recent analysis of the Herschel
far-infrared continuum data of that IRDC has been published by
\citet{pitann2013}. Figure \ref{g48_large} shows the 70\,$\mu$m
Herschel/PACS image outlining the dark features but also several
embedded protostars within the cloud (the area of our carbon spectral
line observations is marked).  Compared to the region within G11.11
discussed in the previous section, G48.66 appears slightly more
evolved, yet still very young.

The integrated emission maps of the different carbon phases of this
region are presented in Figure \ref{g48_overlays}. In G48.66 the
  C$^{18}$O(2--1) and [CI] emission follows the dense gas in east-west
  direction traced by the ATLASGAL 870\,$\mu$m emission, and it
  exhibits an additional extension toward the south of that east-west
  ridge. The [CI] map exhibits even larger differences with more
  emission in east-west extension in the southern part of the map.
These spatial substructures could not be identified by the previous
lower-spatial-resolution [CI] observations by \citet{ossenkopf2011}
but are consistent with the two peaks of emission seen there.

\begin{figure}[htb] 
\includegraphics[width=0.49\textwidth]{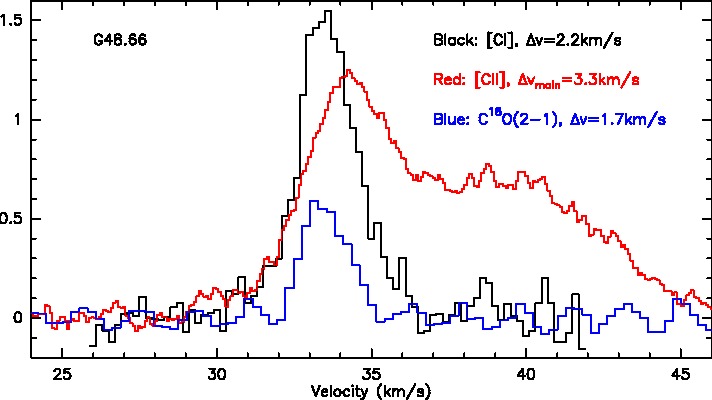}
\caption{G48.66: C$^{18}$O(2-1), [CI] and [CII] spectra (in K)
  averaged over the whole area of emission shown in Figure
  \ref{g48_overlays}. The FWHM of the main peak at 34\,km\,s$^{-1}$
  are presented as well.}
\label{g48_spec}
\end{figure}

\begin{figure}[htb] 
\includegraphics[width=0.49\textwidth]{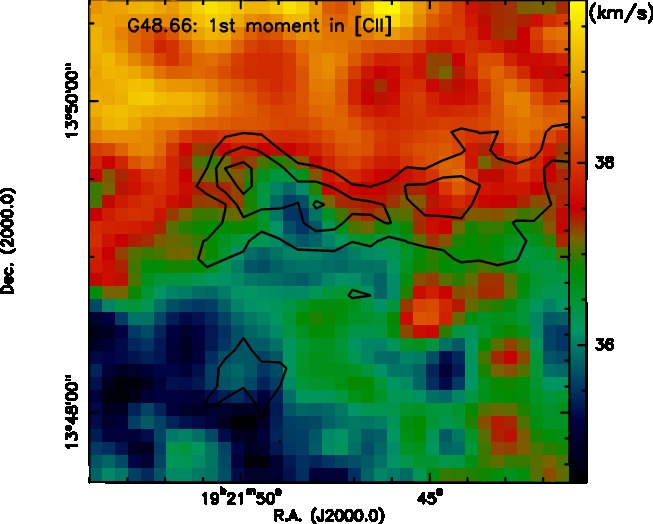}
\caption{G48.66: The color scale presents the 1st moment
  (intensity-weighted velocity) of the [CII] emission between 30 and
  45\,km\,s$^{-1}$.  The contours show the ATLASGAL 870\,$\mu$m emission
  starting at $3\sigma$ level of 150\,mJy\,beam$^{-1}$ and continuing
  in $3\sigma$ steps.}
\label{g48_mom1}
\end{figure}

The biggest difference arises in the ionized carbon [CII] map.  In
contrast to G11.11, where we barely detect the line above the
$3\sigma$ level, here [CII] is strong and shows even different
velocity components. While C$^{18}$O(2--1) and [CI] have single-peaked
spectra around the $v_{\rm{lsr}}$ of $\sim$34\,km\,s$^{-1}$ with
averaged full-width-half-maximum (FWHM) values of 1.7 and
2.2\,km\,s$^{-1}$, respectively (Fig.~\ref{g48_spec}), the [CII]
spectrum exhibits a very broad second component extending up to
45\,km\,s$^{-1}$ (Fig.~\ref{g48_spec}, the FWHM of the main component
at $\sim$34\,km\,s$^{-1}$ is 3.3\,km\,s$^{-1}$). Analyzing the spatial
structure of these two velocity components, we find that the one
around the $v_{\rm{lsr}}$ is similar to the [CI] emission. Although
the [CII] peak position is a bit offset from the C$^{18}$O and dust
continuum peak, the large-scale structure of that component resembles
that of the [CI] and C$^{18}$O emission. In contrast to that, the
high-velocity component is distinctively shifted to the north (the
right two panels in Figure \ref{g48_overlays}). A different way to
dissect the velocity structure of the ionized carbon emission is a 1st
moment map (intensity-weighted peak velocity). The velocity pattern
shown in Figure \ref{g48_mom1} is dominated by a steep velocity
gradient right across the dense cloud filament seen in the submm dust
continuum emission as well as the infrared extinction.  We will
discuss that gradient in more detail in section \ref{flows}.

\subsubsection{IRDC\,18223:} 
\label{18223}

\begin{figure}[htb] 
\includegraphics[width=0.49\textwidth]{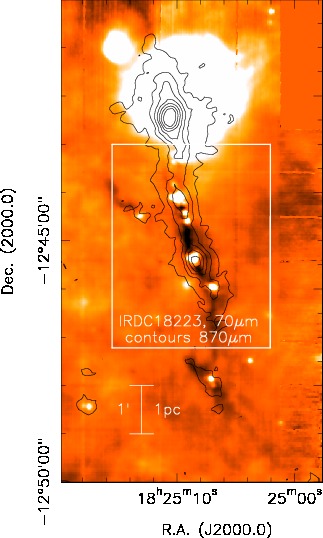}
\caption{IRDC\,18223: Large-scale Herschel/PACS 70\,$\mu$m image of
  the IRDC\,18223 region with 870\,$\mu$m ATLASGAL contours starting
  at 200 and continuing in 300\,mJy\,beam$^{-1}$ steps . The white box
  outlines the region of our carbon observations shown in
  Fig.~\ref{18223_overlays}.}
\label{18223_large}
\end{figure}

The IRDC\,18223 has been studied in detail in recent years on scales
of the filament (e.g., \citealt{beuther2002a,beuther2010b,garay2004})
as well as on smaller scales of individual substructures (e.g.,
\citealt{beuther2007a,fallscheer2009}). This filamentary IRDC hosts
several evolutionary stages from an already evolved high-mass
protostellar object (HMPO) in the north to the dark filament with
embedded very young protostars (Fig.~\ref{18223_large}). It is also
part of a much larger filament extending more than 70\,pc in the Milky
Way plane (e.g., \citealt{tackenberg2013,ragan2014}). For this carbon
study we focus in particular on the infrared dark filament outlined in
Figure \ref{18223_large}.

\begin{figure*}[htb] 
\includegraphics[width=0.99\textwidth]{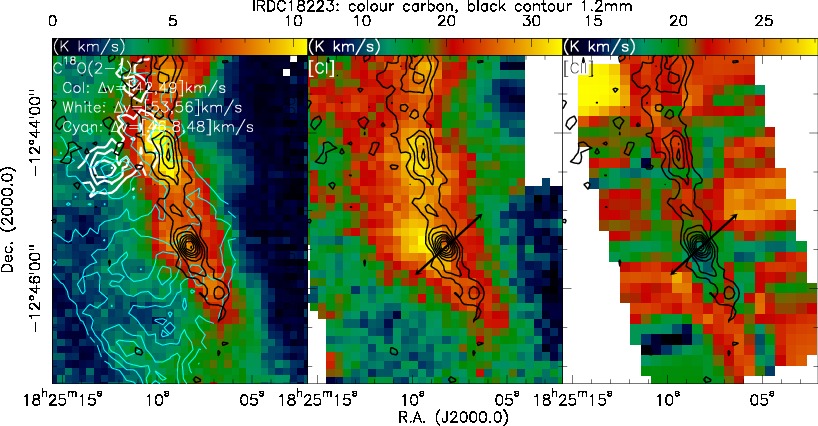}
\caption{IRDC18223: The color scale shows in the left, middle and
  right panel the emission from C$^{18}$O(2--1), [CI] and [CII],
  respectively. The integration regimes for the three color images are
  are [42;49]. [43;49] and [43;55]\,km\,s$^{-1}$. The white and cyan
  contours in the left panel correspond to the velocity components at
  [53.0;56.0] and [46.9;48.0]\,km\,s$^{-1}$. The black contours always
  show the 1.2\,mm continuum emission \citep{beuther2002a} starting at
  $3\sigma$ level of 36\,mJy\,beam$^{-1}$ and continuing in $3\sigma$
  steps. The arrows in the middle and right panel outline the
  direction of a bipolar outflow studied by \citet{fallscheer2009}.}
\label{18223_overlays}
\end{figure*}

\begin{figure}[htb] 
\includegraphics[width=0.49\textwidth]{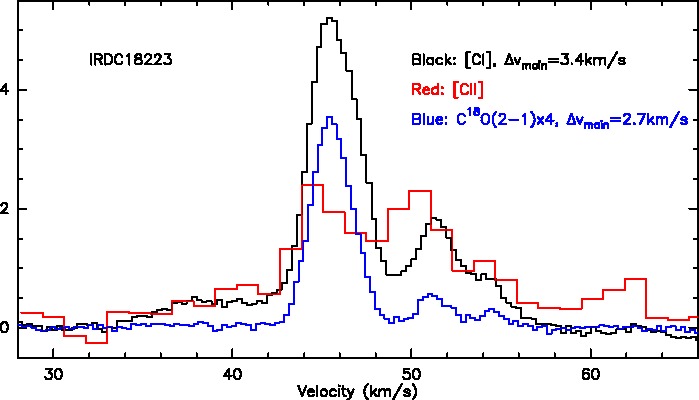}
\caption{IRDC18223: C$^{18}$O(2-1), [CI] and [CII] spectra (in K) averaged
  over the whole area of emission shown in Figure
  \ref{18223_overlays}. The fitted [CI] and C$^{18}$O(2--1) FWHM of
  the main component around 45\,km\,s$^{-1}$ are presented as well
  (class did not converge to a reasonable fit result for the [CII]
  line).}
\label{18223_spec}
\end{figure}

The spatial and spectral structure of the molecular C$^{18}$O,
  atomic and ionized carbon in this region are presented in Figures
  \ref{18223_overlays} and \ref{18223_spec}. Around the $v_{\rm{lsr}}$
  of $\sim$45.5\,km\,s$^{-1}$, the main emission is seen in the
  molecular C$^{18}$O(2--1) and the atomic [CI] emission, and both
  species exhibit a secondary strong peak around 51\,km\,s$^{-1}$. The
  averaged spectrum over the whole region in the ionized [CII] line
  shows emission peaks at similar velocities, however, the intensities
  of both peaks are almost the same. All three species exhibit a third
  spectral peak at slightly higher velocities of
  $\sim$54.5\,km\,s$^{-1}$. The [CI] spectrum in Figure
  \ref{18223_spec} exhibits an additional weak emission component
  between approximately 35 and 41\,km\,s$^{-1}$.  However, the images
  show that this component is only found at the south-eastern edge of
  our map and barely covered at all. Therefore, we do not show it
  here.

\begin{figure*}[htb] 
\includegraphics[width=0.99\textwidth]{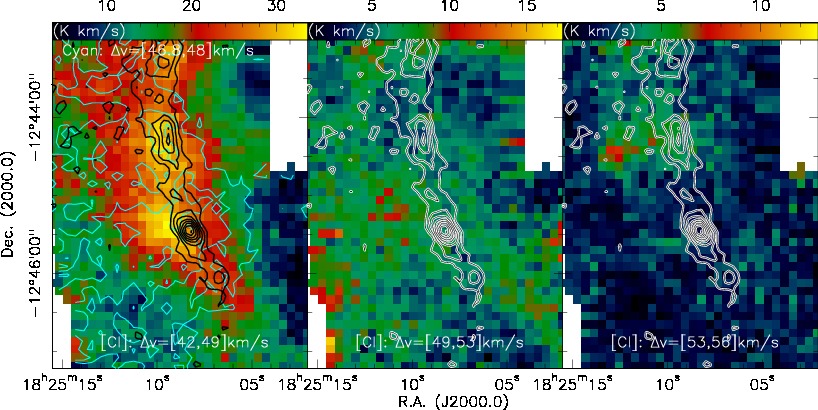}
\caption{IRDC18223: The color scale shows in the left, middle and
  right panel the emission from atomic carbon [CI] in the three
  velocity regimes marked above each panel.  are [42;49]. [43;49] and
  [43;55]\,km\,s$^{-1}$. The cyan contours in the left panel
  correspond to [46.9;48.0]\,km\,s$^{-1}$. The black contours always
  show the 1.2\,mm continuum emission \citep{beuther2002a} starting at
  $3\sigma$ level of 36\,mJy\,beam$^{-1}$ and continuing in $3\sigma$
  steps.}
\label{18223_ci_multi}
\end{figure*}

Figure \ref{18223_ci_multi} shows the spatial structure of these three
velocity components for the atomic [CI] line. The main spectral
component between 42 and 49\,km\,s$^{-1}$ is strongly correlated with
the dense gas traced by the 1.2\,mm dust continuum emission measured
with MAMBO at the IRAM 30\,m telescope \citep{beuther2002a}. However,
already the high-velocity end of this component between 46.8 and
48\,km\,s$^{-1}$ exhibits a spatially distinct structure visible in
the cyan contours in Figure \ref{18223_ci_multi} (left panel). In
contrast to this, the second component between 49 and 53\,km\,s$^{-1}$
which is also strong in the averaged spectrum (Fig.~\ref{18223_spec})
appears spatially diffuse with barely any obvious peak emission
(Fig.~\ref{18223_ci_multi} middle panel). Going to even higher
velocities the spectral component between 53 and 56\,km\,s$^{-1}$ is
spatially very localized with a small peak to the east of the filament
(Fig.~\ref{18223_ci_multi} right panel and Fig.~\ref{18223_overlays}
left panel). The spatial structure of the C$^{18}$O(2--1) emission is
similar to the [CI] line, whereas for the [CII] emission we do not
find clear structures, but the ionized carbon appears to be distributed
much more diffusely over the entire complex without any clear
association with the infrared dark filament. Interestingly, two strong
[CII] emission features lie southeast and northwest of the central
peak in the filament in the direction of the two outflow lobes
discussed in \citet{fallscheer2009}. While this is not conclusive,
there is the possibility that these [CII] emission features could be
partly related to the outflow from the embedded young protostar. 
%The non-detection of [CII] emission toward the central peak in the
%filament (called IRDC\,18223-3) is consistent with a Herschel PACS
%spectrum toward that emission that exhibits a clear absorption profile
%(Linz et al.~private communication).

\subsubsection{IRDC\,18454:} 
\label{18454}

\begin{figure}[htb] 
\includegraphics[width=0.49\textwidth]{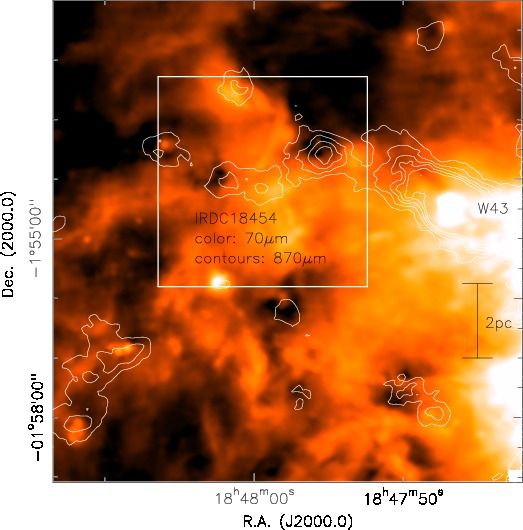}
\caption{IRDC\,18454: Large-scale Herschel/PACS 70\,$\mu$m image of the
  IRDC\,18223 region with 870\,$\mu$m ATLASGAL contours in steps of
  0.4\,mJy\,beam$^{-1}$. The white box outlines the region of our
  carbon observations shown in Fig.~\ref{18454_overlays}.}
\label{18454_large}
\end{figure}

The fourth region in our sample, IRDC\,18454, is in the direct
vicinity of the Galactic mini-starburst W43. Figure \ref{18454_large}
presents the 70\,$\mu$m emission of the region, and while W43 at the
western edge is very bright in the far-infrared, our regions of
interest are either faint at 70\,$\mu$m or they even show dark
absorption patches.

Another intriguing aspect of this region at the interface of the
Galactic bar with the inner spiral arm, is the velocity structure.
Toward this region, all observed spectral lines exhibit 2 spectral
components, one centered at approximately 100\,km\,s$^{-1}$ and the
second at $\sim 50$\,km\,s$^{-1}$ (Figure \ref{18454_spec}). This is
not just observed in the carbon tracers shown here, but also in dense
gas tracers like N$_2$H$^+$ or NH$_3$ (e.g.,
\citealt{nguyen2011,beuther2012a}). Also Galactic surveys of radio
recombination lines show a large fraction of multiple components in
this part of the Galaxy \citep{anderson2011}. These different
components could be caused by two independent clouds in different
spiral arms (e.g., \citealt{nguyen2011}), via colliding gas flows
(e.g., \citealt{carlhoff2013}), or by potential cloud-cloud
interaction in the interface between the Galactic bar and the inner
Scutum spiral arm (e.g., \citealt{beuther2012a}). For a detailed
discussion about this IRDC, the influence of the neighboring W43
mini-starburst and the multiple velocity components see
\citet{beuther2012a}.

\begin{figure*}[htb] 
\includegraphics[width=0.99\textwidth]{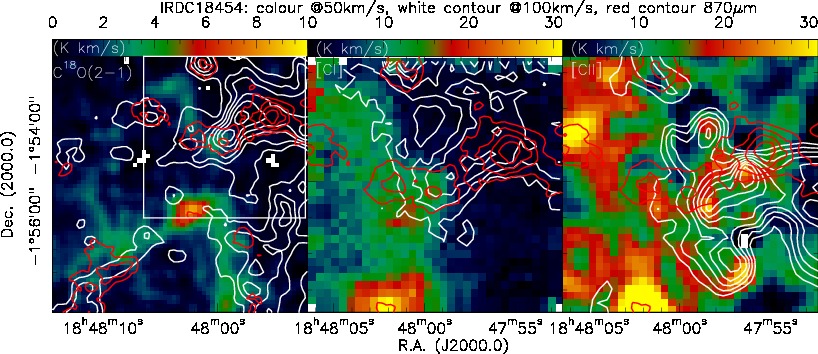}
\caption{IRDC18454: The color scale shows in the left, middle and
  right panel the emission from C$^{18}$O(2--1), [CI] and [CII],
  respectively. The integration regimes for the three color images are
  [47;55]\,km\,s$^{-1}$. The white contours in all three panels
  correspond to the emission from the same species only integrated
  from [80;120]\,km\,s$^{-1}$. The red contours always show the
  ATLASGAL 870\,$\mu$m emission in levels of 0.4\,Jy\,beam$^{-1}$.}
\label{18454_overlays}
\end{figure*}

\begin{figure}[htb] 
\includegraphics[width=0.49\textwidth]{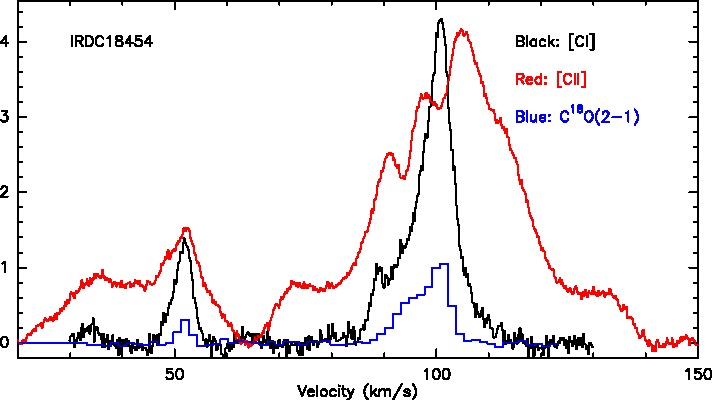}
\caption{IRDC18454: C$^{18}$O(2-1), [CI] and [CII] spectra (in K)
  averaged over the whole area of emission shown in Figure
  \ref{18454_overlays}. }
\label{18454_spec}
\end{figure}

Independent of the interpretation of the different velocity
components, the velocity spread of each individual component is very
broad, for the high-velocity component it ranges from approximately 65
to 140\,km\,s$^{-1}$, and these velocities are all spatially connected
\citep{nguyen2011}. Because of the complex structure of the spectra,
we do not report FWHM values for this region. Figure
\ref{18454_overlays} presents emission of the two components
integrated over a broad part of their spectra. In
\citet{beuther2012a}, it is argued that the two velocity components
appear spatially interacting. In the new carbon data, this is best
visible in the atomic phase where the 50\,km\,s$^{-1}$ occupies the
eastern part and the 100\,km\,s$^{-1}$ mainly the western part of the
observed region. In the ionized carbon [CII], the 50\,km\,s$^{-1}$
emission is comparably very strong, however, the spatial separation of
both components is less obvious. In contrast to that, the
C$^{18}$O(2--1) emission shows in general a similar structure as the
atomic carbon [CI] with the main difference that the 50\,km\,s$^{-1}$
appears less extended in the molecular gas. This is consistent with
the observations of the other sources where we also saw a general
tendency that the molecular C$^{18}$O(2--1) emission is the most
compact.

\subsection{Carbon phases versus hydrogen distributions}
\label{distributions}

In order to quantitatively compare the distribution of the different
carbon phases with the molecular gas of these regions, we constructed dust
temperature and column density maps using the Herschel data obtained
for the {\it Earliest Phases of Star Formation} key program
\citep{ragan2012b}. We used PACS data at 100 and 160\,$\mu$m and SPIRE
data at 250, 350 and 500\,$\mu$m. First all maps were convolved to the
SPIRE 500\,$\mu$m resolution then mapped to identical grids. The
spectral energy distribution (SED) at each pixel was fit employing the
hierarchical Bayesian-fitting algorithm described in \cite{kelly2012},
which makes the assumption that the SED is well-fit by a modified
Planck function as follows:

$$S_{\nu} = \Omega\,N\,\kappa_0\,\Big( \frac{\nu}{\nu_0} \Big)^{\beta}\,B_{\nu}(T_d)$$

where $\Omega$ is the solid angle of the observation, $N$ is the
column density, $B_{\nu}(T)$ is the Planck function, which is
evaluated at the dust temperature, $T_d$, and
$\kappa_0$($\nu$/$\nu_0$)$^{\beta}$ is the dust opacity. We assumed a
$\kappa_0$ of 0.006\,cm$^2$\,g$^{-1}$ at 1.3\,mm wavelengths
\citep{ossenkopf1994}, which includes an assumed dust-to-gas ratio of
150 \citep{draine2011}.  The algorithm fits for $N$, $T_d$ and
$\beta$. For the plots, we also show the equivalent $A_V$ computed
from the relation $N_{H2}$ = 0.95 $\times$
10$^{21}$\,cm$^{-2}$\,($A_V$ / mag).

For the comparison, we used the integrated intensity maps of
C$^{18}$O(2--1), [CI] and [CII] for the four regions with the velocity
regimes presented in Figures \ref{g11_overlays}, \ref{g48_overlays},
\ref{18223_overlays} and \ref{18454_overlays}.  These images were
smoothed to the spatial resolution of the SPIRE 500\,$\mu$m data of
$36.6''$. 

\begin{figure*}[htb] 
\includegraphics[width=0.33\textwidth]{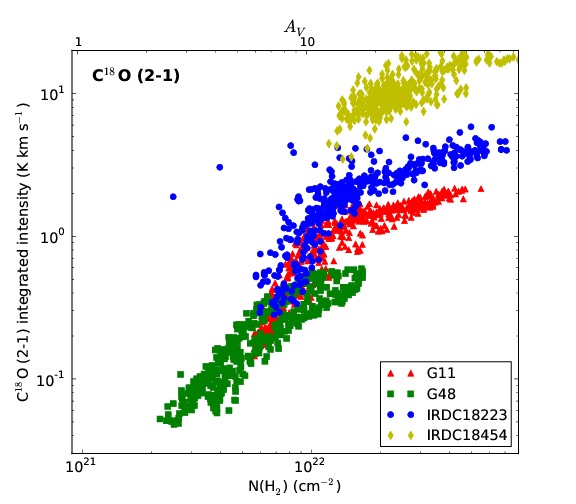}
\includegraphics[width=0.33\textwidth]{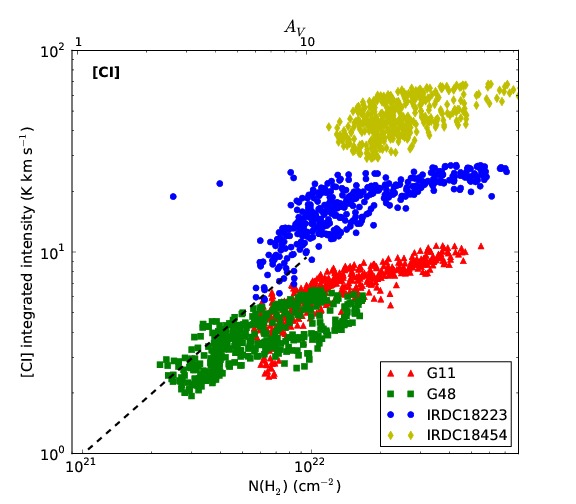}
\includegraphics[width=0.33\textwidth]{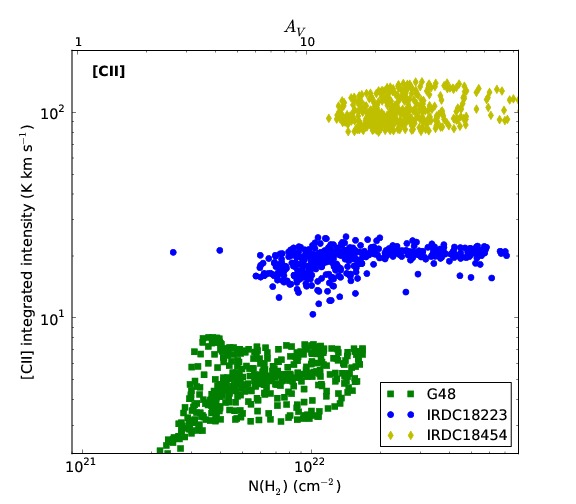}
\caption{Scatter plots of the integrated C$^{18}$O(2--1), [CI] and
  [CII] emission versus the molecular column density and extinction
  (in mag) derived from the Herschel far-infrared data. The
  color-coding separates between the different regions as marked in
  each panel. The integration regimes are the same as those mentioned
  in the previous corresponding figures (for G48.66 only the main
  spectral component and for IRDC\,18454 only the 100\,km\,s$^{-1}$
  component are used). The dashed line in the middle panel shows the
  fit to the modeled [CI] data from \citet{glover2014} who fitted
  their data in their approximately linear regime below 10\,mag
    extinction. For IRDC\,18454, we only use the 100\,km\,s$^{-1}$
  component that is clearly associated with the W43 complex.}
\label{scatter}
\end{figure*}

Figure \ref{scatter} presents the scatter plots of the three gas
phases with respect to the molecular (H$_2$) column density, and one
sees distinct differences between the molecular, neutral atomic and
ionized atomic phases. The C$^{18}$O(2--1) emission strongly increases
at low column densities, and then flattens of at column densities
above $10^{22}$\,cm$^{-1}$.  The latter flattening can likely be
attributed to freeze-out of the carbon monoxide at the low
temperatures in these IRDCs (toward the cloud centers we find
temperatures below 20\,K, see for comparison, e.g.,
\citealt{kramer1999}).

The atomic carbon also shows a correlation with the hydrogen column
density, however, in general the relation is flatter and no clear
break is visible. This confirms the previous morphological assessment
that the atomic carbon is related to the dense molecular gas, however,
weaker than the molecular C$^{18}$O. To compare our results with
the recent cloud formation and [CI] emission models by
\citet{glover2014}, we show as a dashed line in Fig.~\ref{scatter} the
fit they conducted to their model data in their approximately
  linear regime below 10\,mag extinction. The comparison with our
data shows that below column densities of about $10^{22}$\,cm$^{-2}$
(or 10\,mag extinction), the models and observations agree reasonably
well. Toward higher column densities this linear relation between [CI]
emission and molecular column density breaks down and the curves
flatten. This is already indicated in the models by
\citet{glover2014}, and even more prominent in our data since these
observations go to higher column densities.

Finally, the ionized carbon emission exhibits almost no correlation
with the general dense gas distribution.  This confirms that ionized
carbon is distributed in a much broader and widespread component than
the molecular gas.

\subsection{Masses of carbon components}
\label{masses_calc}

In addition to the morphologic analysis, we can use the spectral line
data of the three different carbon phases to estimate the mass
contributions of the ionized and atomic carbon as well as
  molecular CO to the gas phase carbon budget of the interstellar
medium. For all three species, we calculate the corresponding column
densities and masses assuming optically thin emission of the
respective spectral lines.

For the molecular C$^{18}$O, the column densities are calculated using
classical expressions (e.g., \citealt{cabrit1988}):

$$N_{\rm{C^{18}O}} = \frac{6.97\times 10^{15}}{\nu^2 \mu^2} T_{\rm{ex}}e^{E_{\rm{u}}/(kT_{\rm{ex}})} \frac{\tau}{1-e^{-\tau}} \int T_{\rm{mb}}dv ~\rm{[cm^{-2}]}.$$

Here, $\nu$, $\mu$, $T_{\rm{ex}}$, $E_{\rm{u}}/k$, $\tau$ and
$T_{\rm{mb}}$ are the frequency, dipole moment (0.112\,Debye), the
excitation temperature, the upper energy level of 15.8\,K, the optical
depth and the main beam brightness temperature of the line,
respectively.  Assuming $\tau \ll 1 $ the term
$\frac{\tau}{1-e^{-\tau}}$ approaches 1.

The atomic carbon [CI] column densities are calculated following
\citet{frerking1989}:

$$ N_{\rm{[CI]}} = 5.94\times 10^{15} \frac{1+3e^{\frac{-23.6}{T_{\rm{ex}}}}+5e^{\frac{-62.4}{T_{\rm{ex}}}}}{3e^{\frac{-23.6}{T_{\rm{ex}}}}} \int T_{\rm{mb}}dv ~\rm{[cm^{-2}]}$$

and for the ionized carbon [CII], the column densities are calculated following \citet{goldsmith2012}, eq.~26:

$$  N_{\rm{[CII]}} = 3.43\times 10^{16}\left[1+0.5e^{91.25/T_{\rm{kin}}}\left(1+\frac{2.4\times 10^{-6}}{C_{ul}}\right)\right]$$
$$ \times \int T_{\rm{mb}}dv ~\rm{[cm^{-2}]}$$

with the kinetic gas temperature $T_{\rm{kin}}$ and the collision rate
$C_{ul}=R_{ul}n$ depending on the temperature, where $R_{ul}$ is
collision rate coefficient with H$_2$ and $n$ the density.

\begin{figure}[htb] 
\includegraphics[width=0.49\textwidth]{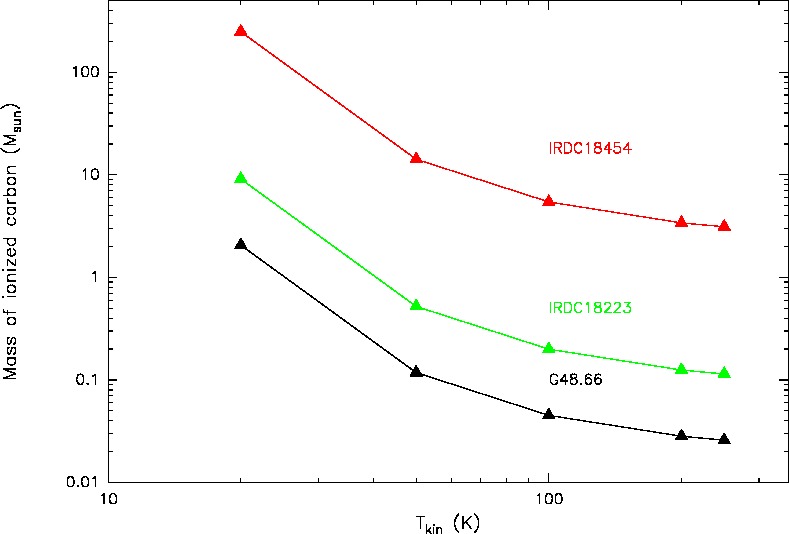}
\caption{Masses of ionized carbon in the three regions where the line
  is clearly detected. We calculated the masses assuming different
  temperatures at an average density of $10^3$\,cm$^{-3}$} as
  outlined in section \ref{masses_calc}.
\label{cii_t}
\end{figure}

The remaining unknown in the equations above is the temperature of the
gas. Since we are dealing with infrared dark clouds, hence the
earliest evolutionary stages, for the molecular C$^{18}$O and the
atomic [CI] we assume 15\,K for all sources following multiple NH$_3$
studies of IRDCs (e.g.,
\citealt{sridharan2005,pillai2006b,dunham2011,ragan2011,wienen2012,chira2013}).
Regarding the temperatures for the ionized carbon, this is less well
determined.  \citet{langer2010} assume temperatures largely between
100 and 150\,K for the diffuse ionized carbon, although above the
level temperature of $E_u/k$ of 91~K, the emissivity is basically
constant, independent of the assumed excitation temperature.  However,
at least in G48.66 and IRDC\,18454 we see a clear spatial correlation
between the ionized carbon emission and that of the dense molecular
CO. This indicates that the temperatures of the ionized carbon could
be lower as well. To estimate the uncertainties based on the
temperature, we calculated the masses of the ionized carbon for
temperatures between 20 and 250\,K at an average density n$\sim
1\times 10^3$\,cm$^{-3}$. The corresponding collision rate
coefficients with H$_2$ are taken from the Leiden database for
molecular spectroscopy (\citealt{schoeier2005},
http://home.strw.leidenuniv.nl/~moldata/) where the para- and ortho
rates are weighted following \citet{lebourlot1991} and
\citet{gerlich1990}. While for IRDC\,18454 in the vicinity of W43
collisions with other partners than H$_2$ may be possible as well, the
good spatial correspondence of the [CII] with the molecular line data
indicates that H$_2$ should be the most dominant collisional partner.
Figure \ref{cii_t} presents the corresponding results. For
temperatures above 100\,K, the corresponding masses do not vary
significantly, however, going to lower temperatures, the values can
vary by more than a factor of a few (see also Table \ref{masses}).

To calculate the masses of the different carbon phases, we integrated
the emission of the different spectral lines over the areas of their
respective emission. For IRDC\,18454, we only use the spectral
component around 100\,km\,s$^{-1}$ since only that is unambiguously
associated with the neighboring W43 complex. Table \ref{masses}
presents the derived masses of the CO (converting C$^{18}$O to
$^{12}$CO masses with a factor 500, \citealt{wilson1994}, applying the
additional CO/H$_2$ ratio of $\sim 10^{-4}$ we have an estimate of
the total gas mass), atomic and ionized carbon in our four target
regions. We find that in all four regions, gas-phase carbon is most
dominant in its molecular form, usually about 10 times more
abundant than atomic carbon.  Interestingly, the ratio between the
ionized and the other two phases is less uniform.  While ionized
carbon is less abundant than atomic carbon in G11.11, we find more
ionized than atomic carbon in the other three regions.  One should
keep in mind that these ratios depend on the covered areas since [CII]
is usually found to be more extended than the other two phases. More
details will be discussed in section \ref{phases}.

\begin{table*}[htb]
\caption{Masses of carbon phases}
\begin{center}
\begin{tabular}{lrrrr}
\hline \hline
phase & G11.11 & G48.66 & IRDC18223 & IRDC18454$^3$\\
      & (M$_{\odot}$) & (M$_{\odot}$) & (M$_{\odot}$) & (M$_{\odot}$) \\
\hline
CO        & 0.81      & 0.30  & 1.84 & 13.4 \\
$[$CI$]$  & 0.056     & 0.025 & 0.21 & 1.6 \\
$[$CII$]$@50K & $<0.012$  & 0.12$^1$  & 0.54 & 14.8 \\
$[$CII$]$@100K & $<0.005$  & 0.05$^1$  & 0.21 & 5.7 \\
CO/$[$CI$]$/$[$CII$]@50K$ & 14.5/1/$>$0.2 & 12/1/4.8 & 8.8/1/2.6 & 8.4/1/3.6$^2$ \\
\hline
Approx. area of emission (pc$^2$)$^4$ & 5.2 & 3.7 & 6.5 & 31.4 \\ 
\hline \hline
\end{tabular}
~\\
{\footnotesize $^1$ The main component between 31 and 37\,km\,s$^{-1}$.\\
$^2$ [CII] calculated at 100\,K because of the energy input from the neighboring W43 region.\\
$^3$ Only the 100\,km\,s$^{-1}$ component is evaluated.\\
$^4$ Based on the [CII] maps.}
\end{center}
\label{masses}
\end{table*}

\subsection{Estimates of bolometric luminosities and radiation fields}
\label{luminosities}

Further interesting parameters to characterize the emission of the
different carbon phases is an evaluation of the bolometric
luminosities in the regions as well as an estimation of the
interstellar radiation field.

The bolometric luminosities are straightforward to estimate. We
extract the total fluxes in our regions from the Herschel far-infrared
data between 70 and 500\,$\mu$m \citep{ragan2012b} and fit spectral
energy distributions (SEDs) to the data points. The resulting SEDs
give us estimates of the bolometric luminosities of the regions.
Table \ref{lum} presents the bolometric luminosities $L_{\rm{bol}}$
derived this way ranging from a few thousand for G48.66 to more than
$10^5$\,L$_{\odot}$ for IRDC\,18454.  These bolometric luminosities
can partly be due to internal sources, but they are also produced by
the external radiation penetrating and heating these clouds. To
approximate the contribution of internal source luminosity to the
total, we use the \citet{ragan2012b} catalog of point sources, which
uses modified blackbody fits to the Herschel point source SED at 70,
100 and 160\,$\mu$m to compute the luminosities. We sum the
luminosities of all point sources in the regions mapped in this work
($L_{\rm{point}}$) and tabulate them in Table \ref{lum}. These values
represent lower limits, as some point sources that were not detected
at all PACS wavelengths are excluded from the Ragan et al. (2012)
catalog, though we expect this additional contribution to be
negligible.  Nevertheless, it is clear that the point source
luminosity contribution is on the order of a few percent of the total
bolometric luminosity of the IRDCs. Thus, we conclude that
$L_{\rm{bol}}$ is dominated by external irradiation.  These bolometric
luminosities may to first order appear large, in particular since we
are studying infrared dark clouds.  However, considering the large
areas over which these bolometric luminosities are derived (between 9
and 31\,pc$^2$, Table \ref{lum}), and taking into account that the
dust emission is also due to external radiation, these values are
plausible. We see a clear differentiation between the more isolated
IRDCs G11.11, G48.66 and IRDC\,18223 with bolometric luminosities in
the $10^4$\,L$_{\odot}$ regime, and the very different IRDC\,18454
that exceeds $10^5$\,L$_{\odot}$. The latter high value is in
agreement with IRDC\,18454 being located in the direct environment of
W43 with an estimated luminosity from the associated Wolf-Rayet
cluster in excess of $10^6$\,L$_{\odot}$ (e.g.,
\citealt{blum1999,beuther2012a}).

It is more difficult to estimate the interstellar radiation field.
However, one can indirectly get an estimate of the interstellar
radiation field from the dust temperature. Here, we follow the
approach outlined in \citet{glover2012}. Assuming that the dust is in
thermal equilibrium, the main process responsible for heating the dust
is the absorption of photons from the interstellar radiation field.
Cooling is dominated by thermal emission from the dust, which for the
dust properties assumed in their study (dust properties of
non-coagulated grains coated with thick ice mantles) scales $\propto
T_{\rm{d}}^6$. In thermal equilibrium, one can then derive a
relation between the interstellar radiation field $G_0$ (here we use
the Draine field which is 1.7 times the Habing field, e.g.,
\citealt{tielens2005}) and the dust temperature $T_{\rm{d}}^6$ as

$$G_0 = \frac{4.7\times 10^{-31}}{5.6\times 10^{-24}} \times \frac{1}{\chi(A_v)} \times T_{\rm{d}}^6$$

where $\chi(A_v)$ is an attenuation factor depending on the extinction.
\citet{glover2012} evaluated $\chi(A_v)$ for the above-mentioned dust
model, and we use $\chi(A_v)\sim 0.1$ corresponding to an average
hydrogen column density $N_H$ of $\sim 10^{22}$\,cm$^{-2}$.

The median dust temperatures were again derived from the SED fits to
the Herschel far-infrared data \citep{ragan2012b} also used in section
\ref{distributions}. The temperatures and corresponding $G_0$ values
are shown in Table \ref{lum}. While the interstellar radiation field
is on the order of a few 10 (in Draine units $=2.7\times
10^{-3}$erg\,cm$^{-2}$s$^{-1}$) for the three more quiescent IRDCs
(G11.11, G48.66, IRDC18223), the estimated external radiation field
exceeds 100 for IRDC\,18454. The latter higher value is again no
surprise since this region is in the direct neighborhood of the
$3.5\times 10^6$\,L$_{\odot}$ mini-starburst W43 (e.g.,
\citealt{beuther2012a}). As obvious in the above equation for $G_0$,
the high sensitivity of the external radiation field on the dust
temperature introduces considerable errors in these estimates.  Since
the four regions have four different temperatures, the values in Table
\ref{lum} give also a rough estimate of the steep dependency of $G_0$
on $T_{\rm{d}}$ and hence the uncertainties associated with this
approach.  Although just ball-park estimates, the luminosities and
radiation fields derived this way are in rough agreement with each
other. The large external radiation field and luminosity derived for
IRDC\,18454 also explain the much stronger emission of [CI] and [CII]
in this complex.

%  Sarah;s email from July 25, 2014
%           T_dust   G_0        L_BOL
%G11           20    54         1.25e4
%G48           21    72         4.83e3
%IRDC18223     18    29         2.15e4
%IRDC18454     23   124         1.94e5
%
% L_point extracted by myself from Sarah's 2012 paper
%       L_point
% G11   16
% G48   136
% 18223 761
% 18454 2269
\begin{table}[htb]
\caption{Luminosities and radiation fields}
\begin{tabular}{lrrrrr}
\hline \hline
Name & $L_{\rm{bol}}$ & $L_{\rm{point}}^a$ & $T_{\rm{d}}^b$ & $G_0$ & area \\
     & ($10^3$L$_{\odot}$) & ($10^3$L$_{\odot}$) & (K) & (2.7E-3$\frac{\rm{erg}}{\rm{cm}^2\rm{s}^1}$)$^c$ & (pc$^2$) \\
\hline
G11.11      & 12.5  & 0.02   & 20 & 54  & 11 \\
G48.66      & 4.8 & 0.14  & 21 & 72  & 9  \\
IRDC\,18223 & 21.5  & 0.76  & 18 & 29  & 13 \\
IRDC\,18454 & 194 & 2.27 & 23 & 124 & 31 \\
\hline \hline
\end{tabular}
{\footnotesize~\\
$^a$ Luminosities of the point sources identified within our fields by \citet{ragan2012b}.\\
$^b$ Median dust temperatures.\\
$^c$ These are commonly referred to as ``Draine units'' (1.7 times the Habing field).}
\label{lum}
\end{table}

%To get an estimate of the external luminosities from the external
%radiation fields, one can multiply the radiation fields by the area we
%have measured the average temperature, hence the areas of our observed
%fields. While this is only a crude approximation, it nevertheless
%gives an approximate estimate of the external luminosities. The this
%way derived external luminosities $L_{\rm{ext}}$ are also shown in
%Table \ref{lum}. Again, we see a clear differentiation between the
%three more quiescent regions G11.11, G48.66 and IRDC18223 that are in
%the range of a few 1000\,L$_{\odot}$ and IRDC\,18454 that
%is exposed to external luminosities in excess of $10^4$\,L$_{\odot}$.
%Since the neighboring Wolf-Rayet cluster has already a luminosity
%$>10^6$\,L$_{\odot}$ (e.g., \citealt{blum1999}), this shows that our
%derived $L_{\rm{ext}}$ values are only rough approximations giving
%ball-park estimates of the environmental luminosities. Nevertheless,
%the environmental differences between IRDC\,18454 on the one hand and
%the other three sources on the other hand are obvious and need to be
%taken into account when interpreting the results.

\section{Discussion}

\subsection{Carbon phases}
\label{phases}

As expected, the C$^{18}$O(2--1) emission traces the dense gas
observable in the dust continuum emission very well. Similarly the
[CI] emission is a reasonable proxy of the dense gas (see also
\citealt{glover2014}), however, it appears somewhat more widespread on
average. The [CII] emission shows a less uniform picture and varies
from a non-detection (G11.11) to compact sources (G48.66 and
IRDC\,18454) as well as more diffuse components (G48.66, IRDC\,18223,
IRDC\,18454). These differences are particularly prominent in the
scatter plots in Figure \ref{scatter}.

Ratios between atomic carbon and CO for typical molecular clouds have
usually been reported in the 20\% to 30\% regime (e.g.,
\citealt{frerking1989,beuther2000}), and \citet{tauber1995} infer a
lower limit of 17\% for the Orion Bar photon dominated region (PDR).
In contrast to that, the largest ratio of atomic carbon to CO we find
is $\sim$12\% in IRDC\,18454 whereas the ratio is lower in the other
regions going down to values of $\sim$7\% in the most quiescent region
G11.11 (\S\ref{sec_sample}). A similar value of 8\% was also found in
the lower-spatial-resolution [CI]/$^{13}$CO study of the G48.66 region
by \citet{ossenkopf2011}. Although our sample of four regions is too
small to derive firm conclusions, we do see a trend of increasing
atomic-to-molecular gas ratio from the earliest, most quiescent
infrared dark and starless clouds to regions that are in a more
evolved environments with radiation sources nearby, e.g.,the high-mass
protostellar object IRAS\,18223-1243 at the northern end of the
IRDC\,18223 or the W43 mini-starburst near IRDC\,18454.

Regarding the ionized carbon, the situation is less conclusive. While
we do not detect [CII] emission in our most quiescent region G11.11,
it is strongly detected in the other regions with higher ionized than
atomic carbon gas abundances. Since the temperatures are more
uncertain for the ionized carbon, the ratios are also less well
determined. For example the ionized-to-atomic gas ratio may vary
between 4 and 9.3 for IRDC\,18454 depending on the assumed temperature
for the ionized carbon of 50 or 100\,K, respectively. The highest
ionized carbon abundances is again seen in the region IRDC\,18454 that is
exposed to the intense radiation field of the W43 mini-starburst.

The Herschel GOTC$^+$ survey has revealed that a significant fraction
of molecular H$_2$ gas is not traced by CO but this sometimes dubbed
``dark gas'' may be traced by the [CII] emission
\citep{langer2010,velusamy2010,pineda2010,pineda2013}. While the
ratio of CO-dark H$_2$ gas depends on the density, \citet{pineda2013}
estimate for the whole GOTC$^+$ survey that approximately 28\% of the
H$_2$ gas can be missed by the CO emission, but these values can go
even above 50\% for the more diffuse lower extinction parts of the
clouds \citep{velusamy2010}. Although deriving CO-dark gas fractions
is not the scope of our paper, we do see wherever we detect [CII]
emission that it is spatially significantly more extended than the
dense C$^{18}$O and dust continuum emission, consistent with the
GOTC$^+$ results. While the C$^{18}$O emission traces the dense gas
associated with the star-forming regions, in particular the [CII]
emission traces the environment and is much more sensitive to the
external UV field. For [CII] emission associated with molecular
clouds, \citet{pineda2010} find that in many cases low far-UV fields
and densities in the regime $10^{3.5}-10^{5.5}$\,cm$^{-3}$ reproduce
the data well, but they also find exception where strong far-UV fields
and higher densities are required. While the latter situation
resembles the IRDC\,18454 complex, the former lower far-UV fields
rather resemble the other three target regions of our mini-survey.

Extended [CII] emission is also regularly observed from strong PDRs, a
recent prominent example being M17SW \citep{perez2012}. They find that
the [CII] emission on the one hand traces parts of the PDR that are
not found by other tracers, and that on the other hand the [CII]
emission is seen from deep in the cloud. The latter is interpreted
as additional evidence of the clumpy nature of the dense ISM. In
contrast to M17SW, we are dealing largely with much lower far-UV
fields, however, again the IRDC\,18454 region partly resembles the
extended emission [CII] emission as well as the association with the
dense gas and hence the clumpiness of star-forming regions.

\subsection{Gas flows}
\label{flows}

Since the different carbon phases trace different parts of the forming
and evolving clouds, the [CII], [CI] and C$^{18}$O lines should harbor
signatures of the cloud formation history or external UV illumination.
In the framework of converging gas flows, kinematic signatures may be
embedded in the more diffuse gas components (e.g.,
\citealt{vazquez2006,heitsch2008,glover2012,clark2012}), and we
investigate such potential signatures in our datasets.

For G11.11 and IRDC\,18454 this turns out to be more difficult. While
in the former source [CII] remains undetected and hence we do not have
much information about the more diffuse parts of the cloud, it is
interesting that we find a line width increase in the C$^{18}$O and
[CI] emission close to a position of multiple N$_2$H$^+$ spectral
features in high-resolution PdBI data (Ragan et al.~in prep.). Such an
enhanced line width or multiple peaks can be interpreted as signatures
of global collapse of high-mass star-forming regions (e.g.,
\citealt{smith2013}).  In comparison to that, in the IRDC\,18454
region the [CII] emission is very strong and shows the multiple
velocity components (\S\ref{18454}). These multiple velocity
components have already previously been reported in molecular and
ionized gas emission \citep{anderson2011,beuther2012a}, and the
controversy stands whether these two velocity components at 50 and
100\,km\,s$^{-1}$ are signs of cloud-cloud interaction at the
galactic-bar/inner-spiral-arm interface or whether they are simply
chance alignments along the line of sight
\citep{nguyen2011,beuther2012a}. This ambiguity may be resolved in the
future when accurate parallax distances from maser sources at both
velocities become available through the Bessel survey
\citep{brunthaler2011}.

Investigating the other two regions, promising kinematic signatures
can be found. For G48.66, in particular the [CII] emission turns out
to be interesting. Although the averaged FWHM line width of the main
gas component in the molecular, atomic and ionized gas does not
exhibit large spreads between 1.7 and 3.3\,km\,s$^{-1}$, the
additional high-velocity gas component visible in Fig.~\ref{g48_spec}
is spatially clearly offset to the north from the main component
around the $v_{\rm{lsr}}$ (Fig.~\ref{g48_overlays}). Looking at the
velocity structure, the first moment map in Fig.~\ref{g48_mom1}
exhibits a strong velocity gradient from north to south directly
across the main infrared dark gas filament and site of active star
formation. This steep velocity gradient at the location of the IRDC
indicates a large velocity shear and by that enhanced Rayleigh-Taylor
instability. Therefore, we suggest that the IRDC may at least
partly be formed through that instability.  Attempts to produce
observational predictions of colliding flow signatures have focused
mainly on CO transitions (e.g., \citealt{heitsch2008,clark2012}), but
the diverse morphologies and dynamics that we observe in the [CII]
line, such as the velocity gradients and correlations with other
tracers described above, give us a promising avenue to potentially
discern between various cloud assembly processes.

The other infrared dark filament IRDC\,18223 does also show different
velocity components, but the kinematic signatures are different.
Comparing only the averaged spectral signatures, the ionized and
atomic carbon as well as C$^{18}$O appear similar with multiple
velocity components (Fig.~\ref{18223_spec}). However, spatially the
ionized carbon is distinct from the atomic and molecular phase, and it
is largely found in a diffuse distribution at the edge of the cloud.
In contrast to that, the multiple velocity components of the atomic
carbon and C$^{18}$O exhibit clear spatial substructures and appear
relatively similar between both phases (Figures \ref{18223_overlays}
and \ref{18223_ci_multi}). The main spectral component around the
$v_{\rm{lsr}}$ of $\sim$45.5\,km\,s$^{-1}$ is clearly associated with
the dense filament traced by the dust continuum emission. But even
within that main component, the more red-shifted part between 46.8 and
48\,km\,s$^{-1}$ is already offset from the filament to the south-east
(Figures \ref{18223_overlays} and \ref{18223_ci_multi}).  Going to
higher velocities, the second spectral component between 49 and
54\,km\,s$^{-1}$ is more diffusely distributed
(Fig.~\ref{18223_ci_multi}), middle panel), whereas at even higher
velocities between 53 and 56\,km\,s$^{-1}$, we find again a compact
gas component east of the filament (Fig.~\ref{18223_ci_multi}) right
panel). As mentioned in section \ref{18223}, the spectrum in
Fig.~\ref{18223_spec} shows an additional blue-shifted [CI] component,
however, that is only found at the lowest edge of the mapped region
and we cannot give more details about that.  With blue- and
red-shifted gas in diffuse as well as compact components found around
the main dense gas filament, we again tentatively interprete these
spectral and spatial structures as evidence that the main filament of
active star formation has formed out of a kinematically active gas
stream that likely converged in the region of the filament.

\section{Conclusions}
\label{conclusion}

With the aim of studying the different carbon gas phases (ionized, atomic
and molecular) during the early high-mass star formation stages, we
observed four infrared dark clouds at high spatial resolution in the
spectral emission lines of ionized ([CII]) and atomic ([CI]) carbon as
well as molecular carbon monoxide (C$^{18}$O(2--1)) with Herschel,
SOFIA, APEX and the IRAM 30\,m telescopes. Except for the ionized [CII]
line in one region, we could map all gas components toward all four
target regions.

\begin{itemize} 

\item While as expected the molecular gas is always closely associated
  with the dense gas filaments visible in infrared extinction as well
  as dust emission, the atomic carbon often has a similar appearance.
  It also traces the dense gas, however, the atomic gas distribution
  appears a bit more extended than the C$^{18}$O maps. The averaged
  line widths within the different regions are slightly smaller for
  the C$^{18}$O(2--1) line compared to the atomic [CI] emission.

\item In contrast to that, the ionized carbon [CII] emission exhibits
  a larger spread in morphologies over the different regions. While it
  remains undetected in the most quiescent region, it shows strong
  emission toward the IRDC in the environment of the W43 starburst.
  Furthermore, the spatial structure of the ionized carbon varies
  between rather diffuse emission at the cloud edges to structures
  that are clearly associated with the dense gas.

\item The data allow us to estimate the relative abundances of the
  different carbon phases within these IRDCs. The ratio between atomic
  carbon and molecular CO mass ranges between 7 and 12\%, lower than
  values found in other molecular clouds, with the lowest values found
  in the most quiescent environment. Although with only four regions,
  the statistical basis is still poor, the data indicate that the
  ratio of atomic to molecular gas depends on the evolutionary stage
  of the region as well as the radiation field the regions are exposed
  to. In the three regions where [CII] is detected, its abundances is
  always higher by a factor of a few than that of [CI].

\item Similar to that, also the ionized carbon is most strongly
  detected in the environment of the W43 mini-starburst. However, for
  the dark clouds the signatures vary significantly. While [CII]
  remains undetected toward the most quiescent region G11.11, it is
  easily and strongly detected toward the G48.66 IRDC that is not much
  more evolved and that also does not have strong radiation sources in
  the environment. Hence, the [CII] emission does not only depend on
  external radiation sources but other factors must be important as
  well.

\item While still in a very young evolutionary stage, G48.66 exhibits
  the strongest kinematic signatures of gas flows in the [CII]
  emission. This indicates that kinematically active regions can
  exhibit stronger [CII] emission as well. Although model predictions
  about kinematic signatures of converging gas flows are scarce or
  non-existent, we interprete the strong [CII] velocity gradient
  across the G48.66 dark filament as suggestive evidence for filament
  formation within kinematically very active gas flows.

\item For the IRDC\,18223, we see multiple red- and blue-shifted
  velocity components. While the spatial structure of the ionized
  carbon in this region is relatively diffuse, the atomic and
  molecular gas exhibits spatially distinct cloud components in the
  environment of the dense filament. Again, we interprete this complex
  spatial and spectral substructures as tentative evidence for gas
  flows that may form the active star formation sites at converging
  points. In addition to this, in IRDC\,18223 we find enhanced [CII]
  emission in the direction of two bipolar outflow lobes. While not
  conclusive yet, it may indicate that [CII] emission can also be
  caused by the kinematic activity of jets and outflows.

\end{itemize}

In summary, mapping star-forming regions at comparable spatial scales
in the ionized, atomic and molecular phases allows us to constrain the
carbon budget as well as the gas kinematics in great detail. Since
cloud and star formation takes place in diverse environments, often
one line does not tell the same story for different regions, hence,
mapping all the different gas phases together is an important mean to
understand the cloud and star formation processes in a global sense.
While this study targeted mainly IRDCs, it will be important to extend
similar studies to more evolved evolutionary stages.

While maps of molecular and atomic gas can be obtained with
ground-based facilities well, Herschel mainly did single-pointing
studies of the ionized carbon, and not many spectral line maps were
obtained with that space observatory.  SOFIA has now proven as a very
efficient observatory to map clouds in the [CII] and other
far-infrared spectral lines, and we hope to get many more exciting
data from this observatory in the future.

\begin{acknowledgements} 
  We would like to thank Rahul Shetty for helping to implement the
  dust temperature fitting algorithm. We also that the anonymous
  referee as well as the editor Malcolm Walmsley for insightful
  comments improving the paper. SER acknowledges support from grant RA
  2158/1-1 which is part of the DFG-SPP 1573 'The Physics of the
  Interstellar Medium'.
\end{acknowledgements}

%\bibliography{/home/beuther/tex/bibliography}   
%\bibliography{/Users/henrikbeuther/tex/bibliography}
%\bibliographystyle{aa}    % this does the style, aa.bst necessary

\end{document}